\documentclass[11pt]{article}

\usepackage{textcomp}  
\usepackage{siunitx}  
\usepackage[version=4]{mhchem}  
\usepackage{amssymb}
\usepackage[utf8]{inputenc}
\usepackage{booktabs}
\usepackage{caption}
\usepackage{lmodern, anyfontsize}
\usepackage[T1]{fontenc}
\usepackage{amsmath,amsthm,bm,mathrsfs,mathtools,hyperref,lipsum}
\usepackage[numbered]{bookmark}
\usepackage{mdframed, comment}
\usepackage{courier, lineno}
\usepackage{textgreek} 
\usepackage{soul}
\usepackage{xcolor}
\usepackage{gensymb}

\usepackage{authblk}
\hypersetup{
    colorlinks,
    citecolor=black,
    filecolor=black,
    linkcolor=black,
    urlcolor=black
}

\usepackage{geometry, float}
\usepackage{graphicx, float, multicol, balance}
\usepackage[version=4]{mhchem}
\usepackage[sorting=none, maxnames=10, style=nature, url=false, backend=biber]{biblatex}
\title{Xenon Anesthesia and Nuclear Spin Effects in Chiral Systems}

\author{

Allan Wang$^{1}$,
S. Furkan Ozturk$^{2}$\thanks{ozturk@caltech.edu},\\

\footnotesize{$^{1}$Division of Chemistry and Chemical Engineering, California Institute of Technology, Pasadena, CA 91125, USA},\\
\footnotesize{$^{2}$Division of Geological and Planetary Sciences, California Institute of Technology, Pasadena, CA 91125, USA}}

\date{\today}

\begin{document}
\maketitle

\begin{abstract}

A general mechanism for anesthetic function is not fully understood. Similarly, the mechanism by which xenon, a chemically inert noble gas, can produce anesthetic effects remains ambiguous. However, a previous study reported a surprisingly strong nuclear-spin–dependent variation in anesthetic potency in mice, although no rigorous molecular mechanism was proposed. This perspective examines that observation and explores a potential connection to the chiral-induced spin selectivity (CISS) effect, a phenomenon that can account for spin-dependent processes in chiral systems. Here we propose a mechanism that links spin-dependent charge organization with chiral molecular systems through a kinetic model that reproduces the reported nuclear spin dependence of xenon anesthesia. The model is based on the nuclear spin–dependent permeability of isotopes through homochiral media, which modulates biological function through ligand–receptor binding in analogy with the Hill–Langmuir equation. Unlike mechanisms that require long-range quantum coherence, our framework remains robust under physiological, room-temperature conditions because it relies on the intrinsic stability of the CISS effect in dissipative biological environments. Our analysis motivates further experimental investigation of spin-dependent processes—not limited to anesthesia—in complex living systems where chirality is pervasive.


\end{abstract}

\section{Introduction}

The mechanism of general anesthesia is a hotly debated topic. While there are empirical trends such as the Meyer-Overton correlation and commonly inhibited or activated receptors, it is generally believed that each class of anesthetics likely recruits a variety of different mechanisms that integrate into a similar network-level phenomenon of communication breakdown \cite{mashour2024anesthesia}. Therefore, it is advantageous to investigate the effect of each anesthetic individually, xenon in our case, without gross generalization.

The inert noble gas xenon is a special anesthetic. It is known as a fast-acting, fast-recovering anesthetic with minimal side effects, albeit with a high cost of production and dependency on specialized equipment for delivery \cite{mcguigan2023cellular}. Current understanding of xenon’s anesthetic mechanism centers on its role as a non-competitive antagonist of the $N$-methyl-D-aspartate (NMDA) receptor. Unlike most volatile anesthetics that primarily modulate $\gamma$-aminobutyric acid type A receptors, xenon is thought to target the glycine-binding site of the NMDA receptor, thereby inhibiting excitatory neurotransmission \cite{dickinson2007competitive}. It also modulates two-pore domain potassium channels and certain nicotinic acetylcholine receptors \cite{mcguigan2023cellular}. These interactions are typically modeled using classical pharmacodynamics, where the anesthetic’s potency is a function of its size, polarizability, and binding affinity to hydrophobic pockets within these proteins \cite{abraini2014crystallographic}. However, the classical paradigm fails to account for a striking observation: the dependence of anesthetic potency on nuclear spin \cite{li2018nuclear}.

\subsection{Anesthetic Potency of Xenon is Nuclear Spin Dependent}

\begin{figure}[H]
    \centering
    \includegraphics[width=0.7\textwidth]{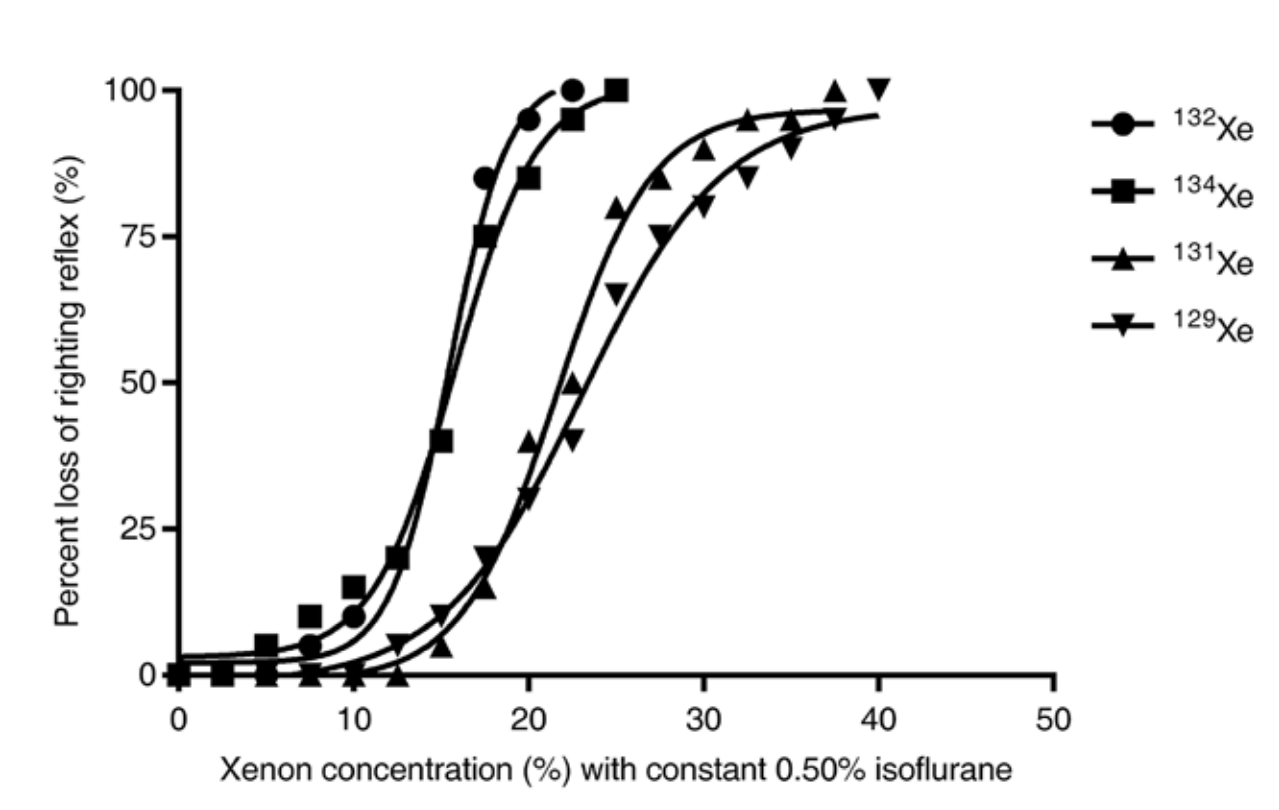}
    \caption{Dose-response curves for xenon isotopes co-administered with 0.50\% isoflurane show reduced anesthetic potency for spin-active isotopes. The x-axis indicates xenon concentration in the test chamber alongside 0.50\% isoflurane. Loss of righting reflex (LORR) is a binary assay scoring whether mice recover upright posture after being placed on their backs within a fixed observation period. Percent LORR represents the fraction of animals (n=20) that failed to recover. The best-fit sigmoidal curves are shown. Spin-active xenon isotopes have a median effective dose (ED50) of 22\%, compared with 15\% for spin-zero isotopes. Reprinted from Li et al. (2018) \cite{li2018nuclear}.}
    \label{Xenon-Mice}
\end{figure}

A recent in-vivo experiment with mice showed that, when co-administered with 0.5\% isoflurane, xenon anesthesia was reported to have a nuclear spin dependent effect \cite{li2018nuclear}. Namely, the xenon isotopes with nuclear spin, $^{129}\text{Xe}$ ($I = 1/2$) and $^{131}\text{Xe}$ ($I = 3/2$), exhibit significantly lower, by about 45\%, anesthetic potency compared to spin-zero isotopes, $^{132}\text{Xe}$ or $^{134}\text{Xe}$ (\autoref{Xenon-Mice}). While the reported signal appears robust, its interpretation, including ours, rests critically on the validity and reproducibility of the underlying observation.

Li et al. did not provide a robust molecular mechanism, instead attributing the phenomenon to the involvement of spin-active xenon in quantum consciousness, suggesting that such processes may partially counteract anesthetic potency \cite{li2018nuclear}. However, this interpretation remains speculative and is not presently grounded in a well-established mechanistic framework.

Given the minimal relative mass differences among these isotopes and their nearly identical electronic structures and chemical properties, conventional mass-dependent kinetics offer no clear explanation for the effect. Moreover, isoflurane is typically administered as a racemic mixture—including in the study by Li et al.—which effectively rules out a nuclear spin–dependent, chirally selective interaction between xenon and isoflurane \cite{li2018nuclear}. Under these conditions, a nuclear spin–dependent interaction of xenon with the biological medium itself remains the leading plausible contributor.

This highly intriguing observation drew our attention because of its potential connection to the recently described spin-dependent phenomenon known as Chiral-Induced Spin Selectivity (CISS). The CISS effect strongly couples electronic spin to chiral molecular systems, even at room temperature and in solution, and has been proposed to influence spin–dependent processes in biological environments, including through hyperfine couplings to the nuclear spin. Investigating this possible link may deepen our understanding of xenon anesthesia and, more broadly, the role of nuclear spin effects in biological systems.

In the following sections, we first critically evaluate two possible mechanisms of xenon anesthesia and its dependence on nuclear spin. We then describe how CISS can account for nuclear spin–dependent effects in biological media. Finally, we introduce a minimal model of anesthetic function in which spin-selective permeability of isotopes in homochiral media is incorporated to account for the observed nuclear spin dependence of xenon anesthesia.


\subsection{Do Xenon Isotopes Have Different Volumes?}

A question in the debate over the mechanism of xenon’s spin-dependent potency is whether isotopic mass differences manifest as variations in atomic volume, thereby influencing membrane permeability or binding within the NMDA receptor's hydrophobic pockets. 

High-precision measurements of highly charged ions via extreme ultraviolet spectroscopy have quantified the difference in mean-square nuclear charge radii between xenon isotopes, such as the $\delta\langle r^2\rangle_{136,124} = 0.269(42) \text{ fm}^2$ shift between $^{136}\text{Xe}$ and $^{124}\text{Xe}$ \cite{silwal2018measuring}. While these fluctuations occur on the femtometer scale—roughly $10^{-5}$ times the diameter of the atom—the Nuclear Field Shift theory suggests that in heavy elements, such nuclear volume changes can slightly perturb the surrounding electron density \cite{bigeleisen1996nuclear}. According to Bigeleisen’s framework, which was applied to heavy uranium isotopes, this nuclear volume change translates into a shift in the electronic distribution; however, the resulting variation in the effective atomic volume is calculated to be on the order of $10^{-4}$ or less. Therefore, even accounting for these field shifts, the resulting changes in atomic polarizability and chemical potential are remarkably subtle, and cannot account for the observed phenomena.


\subsection{The Radical Pair Mechanism}

The radical pair mechanism (RPM) operates as a spin-dependent chemical reaction wherein a pair of spatially separated but spin-correlated electrons oscillates between singlet and triplet states \cite{zadeh2022magnetic}. This interconversion is driven by the hyperfine coupling between the electron spins and nearby magnetic nuclei, as well as by external or internal magnetic fields. In the context of xenon anesthesia, Smith et al. propose that the anesthetic site within a neural receptor—likely the NMDA receptor—facilitates the formation of such a radical pair \cite{smith2021radical}. The central thesis is that the nuclear spin of xenon isotopes acts as a local magnetic perturbation that alters the singlet-triplet yield of these radical pairs compared to spin-zero isotopes, which they assume to affect consciousness.

However, the application of RPM to general anesthesia faces biophysical challenges, primarily concerning the initiation of the radical pair and its downstream effects. In traditional magnetoreception models, such as those involving cryptochrome proteins, radical pairs are generated through photoexcitation. The absence of endogenous photons in the mammalian brain therefore necessitates an alternative mechanism for radical generation. While it is suggested that metabolic processes or reactive oxygen species might serve as a chemical trigger \cite{smith2021radical}, there is currently no evidence for the existence of a thermally activated or metabolically driven radical pair within the specific hydrophobic pockets of anesthetic targets. Furthermore, rapid quantum decoherence—characterized by the $T_2$ time, not the spin relaxation within the $T_1$ time—at physiological temperatures poses an additional challenge, as thermal noise in the cellular environment is expected to collapse a large-scale entangled spin state far more quickly than the millisecond timescales typically required for chemical signaling.

Finally, there is no established mechanistic link between quantum spin dynamics and macroscopic consciousness, even if xenon nuclear spin can influence singlet–triplet yields. The RPM framework remains largely within the quantum-consciousness paradigm, which proposes that the brain operates as a quantum-coherent system with large-scale entanglement. Even if nuclear spin were to modulate the yield of a specific chemical product, no known physiological pathway currently explains how a shift in singlet versus triplet populations could translate into global suppression of the central nervous system.

Having investigated the effect—or lack thereof—of several proposed explanations for the nuclear spin dependence of xenon anesthesia, we now propose a different mechanism that may underlie a broad range of spin-related phenomena in biological systems.

\section{Chiral-Induced Spin Selectivity and Spin Effects in Biological Systems}

The Chiral Induced Spin Selectivity (CISS) effect describes the phenomenon in which molecular chirality determines the preferred spin orientation of electrons transmitted through a system. Although electron spin—a fundamental quantum property—is generally considered decoupled from closed-shell molecules, chiral structures such as DNA and proteins act as highly efficient spin filters, with filtering efficiencies approaching 100\% even at room temperature \cite{bloom2024chiral}. CISS therefore establishes a direct link between molecular chirality and electron spin, making charge transport and electronic reorganization processes in chiral media strongly spin dependent.

A noteworthy advantage of the CISS effect in biological contexts is its inherent robustness at room temperature, contrasting sharply with quantum biological models that rely on long-range entanglement. Although mechanisms such as the RPM depend on long-ranged entangled states—which are prone to rapid decoherence in dissipative biological media—CISS is robust under aqueous-phase, ambient conditions. In fact, heat dissipation is proposed as an ingredient of the CISS mechanism in order to explain its enhancement at higher temperatures \cite{fransson2025chiral}.

In the CISS framework, molecular chirality produces an effective magnetic field in the rest frame of the electron that dynamically measures and breaks the singlet entanglement of an electron pair. Therefore, the requirement for robustness in CISS-based phenomena shifts from phase coherence ($T_2$) to spin relaxation ($T_1$). Given that $T_1$ is typically several orders of magnitude longer than $T_2$, CISS can provide a significantly more feasible pathway for spin-dependent processes to manifest in biological environments.

Among its many manifestations, CISS has been shown to play a significant role in molecular recognition \cite{naaman2022chiral}, heterogeneous interactions between chiral molecules and magnetic surfaces \cite{ben2017magnetization}, redox reactions such as the oxygen reduction reaction \cite{sang2021chirality}, and models on the origin of life’s homochirality \cite{ozturk2023crystallization}. The effect is also likely to be important for biochemical electron-transfer and redox processes in living systems.

The CISS effect may additionally influence the transport properties of isotopes with nuclear spin, since nuclear and electronic spins can couple through hyperfine interactions (HFI), particularly in highly polarizable atoms (e.g., xenon \cite{bounds2023hyperfine}). It is also important to appreciate that HFI reduce the $T_1$ relaxation time of the nuclear spin, introducing a tradeoff between the coupling strength to the electronic spin and the lifetime of the nuclear spin. However, the latter constraint can be circumvented by the continuous dynamic filtering of electron spin within chiral media. In aqueous systems, this interaction is further modulated by the polarization of protons ($H^+$), which can screen electronic charge and spin. When coupled with chiral phonons, these dynamics can drive isotope fractionation—such as the $D/H$ fractionation due to CISS \cite{goren2025coupling}. Although the underlying mechanisms remain under active investigation, recent studies also demonstrate nuclear-spin-dependent transport of oxygen isotopes through aquaporins, highlighting the relevance of nuclear spin in chiral media \cite{vardi2023nuclear}.

Crucially, because CISS couples spin to the molecular frame, its manifestation depends on the net orientation of electronic and nuclear spin degrees of freedom relative to the chiral molecular axes. Biological systems are particularly compatible with this mechanism, as enzymes and bioreceptors provide well-defined binding sites that enforce the necessary alignment and enable short-range interactions. Consequently, the strong interplay between spin degrees of freedom (electronic and, in polarizable systems, nuclear) and molecular chirality should be carefully considered, especially in biological systems composed of homochiral architectures that exhibit strong spin filtering at room temperature and enforce precise molecular orientations.


\section{Spin-Dependent Biological Function in Chiral Media}

We now briefly discuss two likely contributing factors to the mechanism of xenon anesthesia and how spin-dependent interactions due to CISS may influence them\textemdash leading to a magnetic isotope effect. Although our model is agnostic to the underlying molecular structure, the transport equations describing nuclear spin dynamics in chiral media must still be motivated by a biologically relevant framework of xenon anesthesia. In the following sections, we briefly introduce these two contributing factors and then incorporate their implications into the model developed subsequently.

\subsection{Permeability of Xenon in Phospholipid Membranes}


The Meyer–Overton correlation is a well-known empirical relationship linking anesthetic potency to solubility in olive oil. This correlation is often attributed, though not entirely explained, by the requirement that anesthetic compounds permeate cellular lipid membranes or through other amphiphilic structures \cite{campagna2003mechanisms}. Although many exceptions to the rule exist, xenon conforms closely to the predicted relationship. Modern computational studies further support the idea that xenon incorporation into phospholipid bilayers alters membrane fluidity, lateral pressure profiles, and bilayer structure \cite{booker2013biophysical}. Such changes could disrupt membrane protein function in both presynaptic and postsynaptic neurons. Because phospholipids are inherently chiral, it is also possible that xenon solubility within lipid environments depends on nuclear spin. If so, this would provide a direct connection between nuclear-spin effects and lipid-mediated mechanisms of xenon anesthesia. 

To test this hypothesis, we propose measuring the differential solubility of xenon isotopes in oil. A natural xenon gas source provides a mixture of both spin-active and spin-zero isotopes, enabling comparative analysis. Differential incorporation within the membrane could then be quantified by measuring isotope fractionation of the headspace using gas chromatography–mass spectrometry (GC–MS). In addition, vesicles composed of achiral amphiphiles (e.g., fatty acids) can serve as controls. 



\subsection{Binding of Xenon Isotopes to Receptor Proteins}

The most widely accepted mechanism of xenon anesthesia involves its binding to and inhibition of the NR1 subunit of the NMDA receptor \cite{franks1998how, dickinson2007competitive}. Molecular modeling and inhibition assays have shown that xenon binds competitively at the glycine co-agonist binding site while acting noncompetitively with respect to NMDA itself \cite{dickinson2007competitive, mcguigan2023cellular}. To test whether xenon binding to the NMDA receptor depends on nuclear spin, similar inhibition assays could be performed using isotopically pure xenon samples with and without nuclear spin. 

As noted by Li et al., the inclusion of 0.5\% isoflurane may also influence the results \cite{li2018nuclear}. Isoflurane is known to inhibit NMDA receptors competitively at the glycine co-agonist binding site \cite{dickinson2007competitive}. Xenon is similar in size to glycine, whereas isoflurane is substantially larger. Because up to three xenon atoms may occupy the glycine binding site \cite{armstrong2012identification}, one or two xenon atoms could competitively inhibit isoflurane binding while still leaving sufficient space for glycine-mediated activation of NMDA receptors. Consequently, low concentrations of xenon may reduce the anesthetic potency of isoflurane.




\section{Kinetic Model}

\begin{figure} [H]
    \centering
    \includegraphics[width=0.8\linewidth]{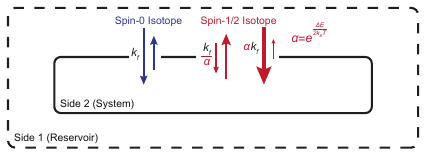}
    \caption{Illustration of the spin-dependent transport kinetic model. The model consists of a reservoir (Side 1) and a finite-volume system (Side 2), where anesthetic action takes place. The system–reservoir interface is chiral. The system (Side 2) and the chiral interface together can be viewed as the active site and the opening or tunnel of a receptor protein, respectively. Transport of xenon across the chiral system–reservoir interface is shown by arrows, with sizes corresponding to the effective rate constants ($k_f$ and $k_b$). Spin-0 isotopes are shown in blue, with $k_f > k_b$. Spin-1/2 isotopes are shown in red, with their two nuclear spin states interacting with the chiral interface. When forward transport of one spin state (e.g., $+1/2$) is favored by the CISS effect, the corresponding forward rate constant $k_f$ is enhanced by a factor of $\alpha$, while the reverse rate constant $k_b$ for the same spin state is reduced by the same factor $\alpha$. The opposite modulation applies to the other spin state (e.g., $-1/2$).}
    \label{fig:2 illustration}
\end{figure}

Both the lipid-based Meyer–Overton framework and the NMDA receptor binding model involve xenon passing through chiral interfaces and interacting with homochiral biomolecules. Our kinetic model is intended to capture the spin-dependence of these interactions due to CISS.

The transport of xenon isotopes across a chiral interface is modeled as a two-compartment system: Side 1 and Side 2. $N_{1,\sigma}$ represents the concentration of xenon with spin $\sigma$ in the input compartment (Side 1) and $N_{2,\sigma}$ represents the concentration of xenon with spin $\sigma$ in the active compartment (Side 2), where anesthetic function occurs upon binding to an active site.

Our model describes two processes. The first is the nuclear-spin dependent transport of xenon isotopes through a chiral medium into a compartment of finite volume. The second is the binding of xenon isotopes that have been transported to the active region of a receptor with a limited number of binding sites, in analogy with the Hill-Langmuir adsorption model. As such, the first process accounts for spin-dependent filling of a compartment, while the second connects this to biochemical potency. 

To maintain generality, we do not assign a particular biomolecular structure for the chiral medium, which may consist of a combination of relevant biological components such as phospholipids and membrane proteins. Our only assumption is that the medium is homochiral and enables spin-dependent transport within the \textit{molecular frame} of the chiral molecule due to CISS.

\subsection{Spin-Dependent Transport}

The exchange of xenon atoms between two compartments can be modeled using first-order transport kinetics:

\begin{equation}
N_{1,\sigma} \overset{\displaystyle{k_f}}{\underset{\displaystyle{k_b}}{\rightleftharpoons}} N_{2,\sigma}
\end{equation}

where $k_f$ and $k_b$ are the forward and backward rate constants, respectively. Within the grand canonical ensemble, with a constant chemical potential for Side 2, $\mu_2$, we define a crowding factor $\beta$ that slows down the forward rate and enhances the backward rate as Side 2 becomes more populated. This factor is defined as $\beta \equiv e^\frac{N_2\mu_2}{k_BT}$, where $k_B$ is Boltzmann's constant and $T$ is the temperature. Assuming Side 1 acts as a large reservoir to Side 2, such that $\mu_2\gg \mu_1$, crowding on Side 1 is negligible. Consequently, the rate constants become number-dependent: $k_f \rightarrow k_f/\beta$ and $k_b \rightarrow k_b\beta$.

The temporal evolution of the system is described by a set of coupled ordinary differential equations (ODEs). The spin-zero xenon isotope's concentration in each compartment evolves as:

\begin{equation}\label{ODE1}
    \frac{dN_{1,0}}{dt}  = -\frac{dN_{2,0}}{dt} = -k_{f}N_{1,0} + k_{b} N_{2,0}
\end{equation}


To describe the dynamics of spin-active isotopes, we modify the transport rate such that the permeability—or the binding efficiency—is sensitive to the nuclear spin state $\sigma \in \{\uparrow, \downarrow\}$, via the substitution $k\rightarrow\alpha^{\pm1} k$. The chiral enhancement factor $\alpha$ is defined through a Boltzmann factor, where $\pm\frac{1}{2}\Delta E$ is the spin-dependent energy shift of up-spin and down-spin isotopes interacting with a homochiral medium due to the CISS effect:

\begin{equation}
    \alpha \equiv \exp\left( \frac{\Delta E}{2k_B T} \right)
\end{equation}

This spin-chirality factor makes the flux equations asymmetric for spin-active isotopes, favoring one spin state while penalizing the other due to the spin-dependent interaction with the chiral medium (Figure~\ref{fig:2 illustration}). 

Incorporating the spin-chirality factor $\alpha$ and the longitudinal nuclear spin relaxation times in compartments 1 and 2, $T_{1,1}$ and $T_{1,2}$ respectively, the concentrations of each isotope $\sigma$ in compartments 1 and 2, $N_{1,\sigma}$ and $N_{2,\sigma}$, evolve as:
\begin{subequations}
\label{eq:ODE2}
\begin{align}
    \frac{dN_{1,\sigma}}{dt} &= -k_f \alpha^{\sigma} N_{1,\sigma} + k_b \alpha^{-\sigma} N_{2,\sigma} - \sigma\frac{N_{1,\uparrow} - N_{1,\downarrow}}{2T_{1,1}} \label{eq:ODE2a} \\
    \frac{dN_{2,\sigma}}{dt} &= +k_f \alpha^{\sigma} N_{1,\sigma} - k_b \alpha^{-\sigma} N_{2,\sigma} - \sigma\frac{N_{2,\uparrow} - N_{2,\downarrow}}{2T_{1,2}} \label{eq:ODE2b}
\end{align}
\end{subequations}

where $\sigma = +1$ corresponds to the spin-up ($\uparrow$) state and $\sigma = -1$ to the spin-down ($\downarrow$) state of the spin-active xenon isotope. The first two terms in \autoref{eq:ODE2a} and \autoref{eq:ODE2b} describe chirality-weighted transport: the forward rate $k_f$ is enhanced (suppressed) by $\alpha$ for spin-up (spin-down) xenon, and the backward rate $k_b$ is suppressed (enhanced) respectively. The third term describes longitudinal spin relaxation within each compartment.

For $\sigma = 0$, the transport kinetics of the spin-zero isotope (\autoref{ODE1}) are recovered. Note also that $d(N_{1,0} + N_{2,0})/dt = d(N_{1,\uparrow} + N_{1,\downarrow} + N_{2,\uparrow} + N_{2,\downarrow})/dt = 0$, confirming that the total concentration of each xenon isotope remains constant. 

\subsection{Using Hill Equation to Model Anesthetic Potency}

The first part of our model predicts the time-dependent concentrations of xenon isotopes in the active compartment. We then model how these concentrations regulate biochemical function in the binding site, using the equilibrium concentration of xenon in Side 2 as the input parameter: $[Xe]=\sum_{\sigma}N_{2,{\sigma}}(t\rightarrow\infty)$ 

Independent of the specific macromolecular target, the sigmoidal dependence of anesthetic potency on xenon concentration is consistent with cooperative ligand-receptor binding, and is well described by the Hill equation. We therefore model the fractional occupancy of the receptor binding site as a function of the concentration of xenon ligands in the active compartment and assume that the anesthetic effect is proportional to this occupancy. 

Specifically, the biological effect arises from the reversible binding of xenon atoms to a functional site $S$ to form an active complex $XeS$:

\begin{equation}
    n[Xe] + [S] \rightleftharpoons [Xe_nS]
\end{equation}

Here $n$ describes the cooperative action of xenon atoms needed to unlock the biochemical function, which is related to the Hill coefficient in the Hill equation.

At equilibrium, the dissociation constant $K_d$ is given by:

\begin{equation}
    K_d = \frac{[Xe]^n [S]}{[Xe_nS]} \implies \frac{[Xe_nS]}{[S]} = \frac{[Xe]^n}{K_d}
\end{equation}

To quantify the anesthetic potency of xenon, we use the fractional occupancy, $\theta$, which is defined as the ratio of occupied sites to total sites $[S]_{tot} = [S] + [Xe_nS]$:

\begin{equation}\label{Hill}
    \text{Potency} \Leftrightarrow \theta \equiv \frac{[Xe_nS]}{[S] + [Xe_nS]} = \frac{[Xe]^n}{[Xe]^n + K_d}
\end{equation}

The $n = 1$ special case of \autoref{Hill} is the Monod equation describing the growth of microorganisms in an environment with limited nutrients. For $n > 1$, \autoref{Hill} describes a cooperative mechanism for xenon anesthesia, yielding a sigmoidal response centered at $[Xe] = K_d$, where the anesthetic potency is exactly $50\%$. 

\section{Results} 

In our simulations, we used $T=310K=36.85\degree C$, $T_{1,1}=10s$, and $T_{1,2}=1s$, based on empirical $T_1$ relaxation time data for hyperpolarized $^{129}\text{Xe}$ ($I = 1/2$) nuclei $T_1$ in oxygenated blood and brain tissues \cite{albert1999t1, choquet2003method}. The longer $T_1$ value in blood (Side 1 in our model) compared to the value in the cell (Side 2 in our model) is attributed to the homogeneous extracellular environment, while intracellular environments accelerate spin relaxation due to molecular interactions, confinement effects, and coupling to paramagnetic species such as reactive oxygen species \cite{saam2015t1}.

\subsection{Transport Kinetics}

\begin{figure}[H]
    \centering
    \includegraphics[width=0.95\textwidth]{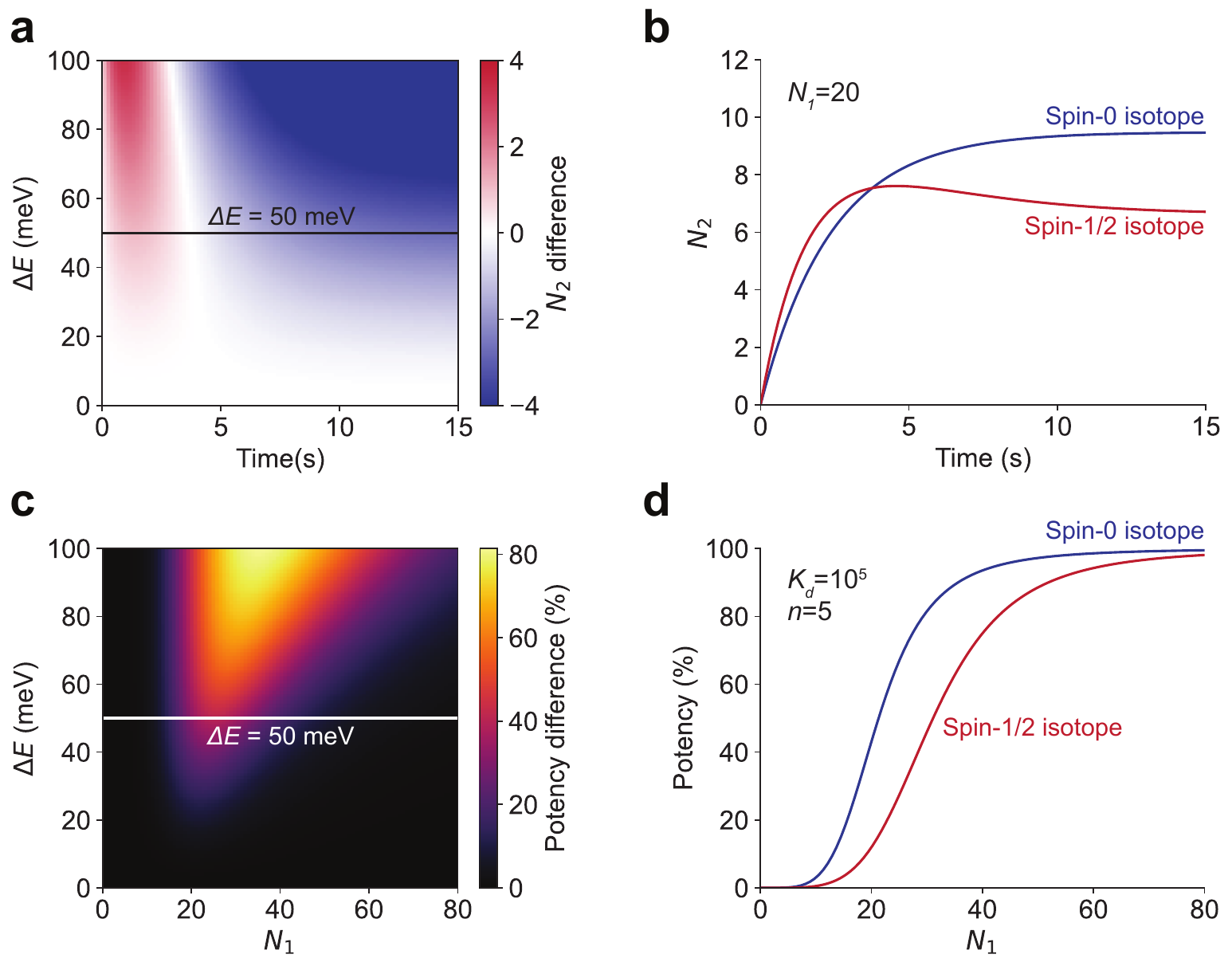}
    \caption{Spin-selective permeability model reproducing nuclear spin-dependent anesthesia of xenon. Top panels 
    (\textbf{a},\textbf{b}) show time-dependent solutions of the spin-dependent transport kinetics; bottom panels (\textbf{c},\textbf{d}) map equilibrium concentrations in Side 2 to anesthetic potency via the Hill equation. 
    \textbf{a} Heat map of the concentration difference between spin-1/2 and spin-0 isotopes as a function of $\Delta E$ and time, with initial condition $N_1=20$. Red indicates higher $N_2$ for spin-1/2, blue indicates higher $N_2$ for spin-0. The horizontal slice at $\Delta E = 50,\mathrm{meV}$ is shown in \textbf{b}. \textbf{b} Time evolution of Side 2 concentrations ($N_2$) at $\Delta E = 50,\mathrm{meV}$, showing initial dominance of the spin-active isotope followed by a spin-0–dominated equilibrium. \textbf{c} Heat map of potency difference between spin-0 and spin-1/2 isotopes as a function of $\Delta E$ and input concentration $N_1$, computed using the Hill equation ($n=5$, $K_d=10^5$). The color bar shows the percent potency difference between isotopes with positive values corresponding to higher potency for the spin-0 isotope. The horizontal line at $\Delta E = 50,\mathrm{meV}$ is shown in \textbf{d}. \textbf{d} Potency comparison at $\Delta E = 50,\mathrm{meV}$, reproducing the spin-dependent anesthetic effect of xenon reported by Li et al., \cite{li2018nuclear}.} 
    \label{data}
\end{figure}

First we investigated how spin-dependent transport affects the equilibrium concentration of $N_2$ in \autoref{data}a and \autoref{data}b. For these simulations, the initial condition $N_1 = 20$ and the baseline rates $k_f=0.20s^{-1}$ and $k_b=0.175s^{-1}$ were used. The model assumed an enhanced affinity of xenon for Side 2 relative to Side 1, motivated by the higher solubility of xenon in neuronal cells than in plasma \cite{chen1980tissueblood}.

As Side 2 became increasingly populated, the crowding factor, $\beta=e^{N_2/(k_BT/\mu_2)}$, based on the chemical potential of Side 2, was dynamically adjusted and modulated both the forward and backward transport rates. We chose $k_BT/\mu_2 = 80$, where this value represents the number of xenon atoms in Side 2 at which crowding effects approached equilibrium.

The figure of merit for spin-dependent effects was determined by $\Delta E$, which is expected to lie in the range of several tens to hundreds of meV based on previous CISS measurements \cite{safari2024enantioselective, bloom2024chiral}. We selected a conservative estimate of $\Delta E=50$meV. At body temperature, $310K$, this gave $\alpha = e^{\frac{50meV}{2k_BT}}= 2.55$. Calculations over a range of $\Delta E$ values were also performed to investigate the overall strength of the effect as a function of this free model parameter, which captured the spin dependence due to CISS.

All initial value problems began with all xenon molecules in Side 1 ($N_1=20$) and $N_2 = 0$. The system was then evolved stepwise according to the differential equations listed in Section 4.1. The time-evolved solutions for xenon transport comparing spin-0 and spin-1/2 isotopes were shown in \autoref{data}b for $\Delta E=50$meV. Initially, the favored spin population of spin-1/2 nuclei moved into Side 2 at a faster rate than spin-0 xenon. As the favored spin population of spin-1/2 nuclei in Side 1 became depleted, the spin-1/2 isotope transitioned to a slower transport regime at approximately $t=3s$. This crossover was sensitive to the choice of $T_1$ relaxation values, and the opposite behavior would have been observed for the non-physiological case of longer spin relaxation time at the active site, as shown in SFigure 1. Eventually, the spin-0 isotope surpassed the spin-1/2 isotope and maintained a higher concentration, $N_2$, at equilibrium. The same initial value problems were evaluated over a range of different $\Delta E$ values in \autoref{data}a, showing that the equilibrium difference in $N_2$ was strongly dependent on $\Delta E$. As expected, larger $\Delta E$ values produced a greater enhancement in the spin-dependent equilibrium concentration difference, and the difference practically vanished below $\Delta E=25$meV.

\subsection{Anesthetic Potency}

The potency of each isotope was investigated in \autoref{data}c and \autoref{data}d. The initial value problems shown in \autoref{data}a were solved over a range of starting concentrations, from $N_1=0$ to $N_1=80$. The resulting equilibrium concentration of nuclei in Side 2, $N_2$, was then used to calculate anesthetic potency using the Hill equation (\autoref{Hill}). We chose $n=5$ as the Hill coefficient, which is the low-end value of a cooperativity range that is consistent with in-vivo anesthesia experiments \cite{eger2001relevant}. A value of $K_d=10^5$ was found to qualitatively reproduce the sigmoidal dose–response relationship shown in Figure 1. 


The resulting dose–response curve for $\Delta E=50meV$ is shown in \autoref{data}d. Notably, the model recapitulates the spin-dependent anesthetic effect of xenon measured by Li et al., shown in \autoref{Xenon-Mice}. Because the equilibrium concentration in Side 2 is lower for the spin-$1/2$ isotope, the functional response is shifted relative to the spin-0 isotope. For the parameters shown, the ED50 value increases by approximately 50\%, consistent with the experimental observations. As with the transport kinetics, the spin-dependent potency difference was calculated over a range of $\Delta E$ values in \autoref{data}c. As expected, the potency difference vanishes below $\Delta E=20$meV.

\section{Discussion}




A primary strength of our minimal model is its ability to probe spin-dependent transport of isotopes within chiral environments without requiring detailed knowledge of specific molecular docking sites. By focusing on asymmetric filtering of spin states during transport, the model reproduces observed variations in isotope concentrations, which are then associated with differences in anesthetic potency between spin-active and spin-0 isotopes. This level of description provides a coarse-grained, transport-based framework for interpreting nuclear spin-dependent biological effects. A consequence of this structural abstraction is that the model does not explicitly incorporate the geometry of the NMDA receptor or other putative binding sites. As a result, it does not resolve or distinguish between specific molecular mechanisms operating after the gas reaches its functional target. Instead, it assumes that downstream biological response can be effectively parameterized as a function of the spin-dependent distribution established before the binding event. In this sense, the model should be understood as addressing an upstream transport-filtering stage rather than the receptor-level mechanism of xenon anesthesia and its nuclear spin dependence.

It is also important to clarify the scope of the framework with respect to the xenon anesthesia problem. The present model does not claim to resolve the long-standing question of the full anesthetic mechanism of xenon, which may involve multiple interacting pathways and cooperative effects. Rather, it isolates and analyzes a pre-binding contribution, in which spin selectivity during transport through chiral biological media (such as phospholipid bilayers or protein pores) can generate isotope-dependent concentration differences at the active site. In this view, anesthesia provides a macroscopic observable that may reflect, in part, earlier spin-dependent transport processes.

The sensitivity of our results to longitudinal relaxation time ($T_1$) highlights an intriguing concept in quantum biology: the ability to modulate biological function by tuning the $T_1$ time of spin-active species in a medium. According to our model, the $T_1$ time is shorter within the functional site, which penalizes the spin-active isotopes. However, for the opposite case of longer $T_1$ time in Side 2 compared to Side 1, spin-active species outcompete the spin-0 ones for all times, including the equilibrium (SFigure 1). The outcome of this causes an inverted biological response whereby spin-active isotopes are more potent anesthetics (SFigure 1d). This consequence of our model indicates that $T_1$ is not merely a spectroscopic parameter but a functional knob that could potentially be manipulated—perhaps via local magnetic environments or local viscosity manipulations—to control biological outcomes. While speculative in terms of direct biological control, this result motivates considering spin relaxation time as a potentially relevant factor to control biochemical reactivity.

Xenon provides a particularly useful system to explore these ideas experimentally. Its chemical inertness isolates isotope effects from conventional chemical reactivity, while its high polarizability makes it sensitive to magnetic interactions within biological structures. In addition, the existence of both spin-active isotopes ($^{129}$Xe, $^{131}$Xe) and spin-0 isotopes ($^{132}$Xe, $^{134}$Xe) enables controlled comparisons that can, in principle, separate nuclear spin effects from mass-dependent contributions.

More broadly, CISS has primarily been studied in isolated molecular and condensed-matter systems. The present work suggests that analogous spin-selective transport processes may also arise in homochiral biological environments. While the results do not directly demonstrate an in vivo manifestation of the CISS effect, they are consistent with the possibility that spin-dependent transport through biological media could lead to novel magnetic isotope effects. If validated experimentally, such a mechanism would provide a route for connecting spin-dependent transport with macroscopic biological responses, including anesthesia, in complex living systems.

\section{Conclusion}

Motivated by experimental observations of nuclear-spin dependence in xenon anesthesia, we developed a model for spin-dependent nuclear transport in chiral biological media due to the chiral-induced spin selectivity effect, a robust phenomenon in dissipative biological environments. Without invoking detailed receptor-level mechanisms or relying on unfeasible long-range spin coherence, our model demonstrates how spin-selective transport prior to target binding can produce substantial differences in anesthetic potency, qualitatively consistent with the observations of Li et al. \cite{li2018nuclear}. Although the present work does not establish the full mechanism underlying xenon anesthesia, it identifies a physically transparent pathway through which nuclear spin and molecular chirality may influence biological function. These findings motivate further experimental investigation of spin-dependent processes in complex living systems that may arise from the chiral-induced spin selectivity effect.

\section*{Acknowledgments}

The authors thank Ron Naaman, Noam Lotem, and John Eiler for fruitful discussions and insightful suggestions on the manuscript. S.F.O. acknowledges support from the Caltech startup funds and the William H. Hurt Scholarship Program.

\printbibliography

\end{document}